# On the Path to High Precise IP Geolocation: A Self-Optimizing Model

Peter Hillmann    Lars Stiemert    Gabi Dreo    Oliver Rose

*Research Center CODE (Cyber Defence),*
*Universität der Bundeswehr München,*
*Werner-Heisenberg-Weg 39, 85577 Neubiberg, Germany*
*{peter.hillmann, lars.stiemert, gabi.dreo, oliver.rose}@unibw.de*

## Abstract

*IP Geolocation is a key enabler for the Future Internet to provide geographical location information for application services. For example, this data is used by Content Delivery Networks to assign users to mirror servers, which are close by, hence providing enhanced traffic management. It is still a challenging task to obtain precise and stable location information, whereas proper results are only achieved by the use of active latency measurements. This paper presents an advanced approach for an accurate and self-optimizing model for location determination, including identification of optimized Landmark positions, which are used for probing. Moreover, the selection of correlated data and the estimated target location requires a sophisticated strategy to identify the correct position. We present an improved approximation of network distances of usually unknown TIER infrastructures using the road network. Our concept is evaluated under real-world conditions focusing Europe.*

## 1. Introduction

*Geolocation* describes the process of allocating a real-world location, e.g. defined by longitude, and latitude, to a virtual address. Determining the geographical location of a network entity is called *IP Geolocation* by using the Internet Protocol (IP) address [1]. Figure 1 shows an example of measurement based IP localization.

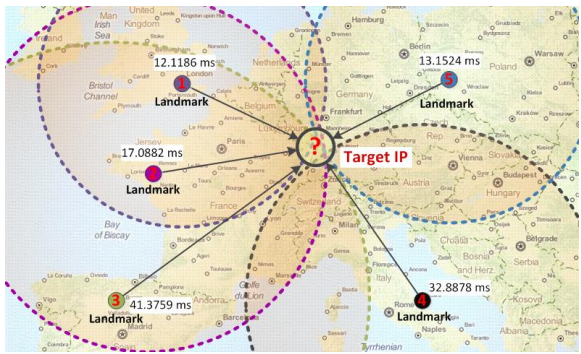

Figure 1. Example of measurement based IP localization.

In times of *Ubiquitous Computing* and *Internet of Things* interconnected information and communication technologies are penetrating the daily life. More and more applications are taking into account from where users are accessing a service. Thus, increasing interest in Geolocation strategies is not only shown by academic research facilities. Also government authorities as well as commercial enterprises benefit from research and technological development in this field. Among others, important use cases are targeted advertising [2], Content Delivery Networks (CDNs), e.g. in terms of optimized load balancing and traffic management [3] as well as cyber warfare. Especially for law enforcement agencies a highly accurate location determination is mandatory in order to successfully prosecute an attacker and fulfill all demands of a forensic strategy. The knowledge of the origination of an attack is mandatory to be able to trace back an attacker and to determine which legal authority is in charge. Therefore, the necessity for a highly precise and reliable IP Geolocation service as well as deviation estimation has been identified as an important goal for the Future Internet [4]. According to this, achieving accurate results requires the consideration of real-world influences in today's network infrastructures.

This work introduce an optimized approach for advanced modeling of IP Geolocation by using latency measurements. It uses the moderate correlation between network delay and geographic distance. Nevertheless, the accuracy is mainly influenced by the selection of the measurement points and the mathematical modelling. Based on a pre-defined set of known hosts, called *Landmarks* or *Vantage Points*, and public available data of the TIER 1 infrastructure, our algorithm identifies optimized positions of measurement points in respect to the overall network. Various conducted measurements serve as input to our model. With help of the introduced more accurate mathematical model and self-optimizing approach, we are able to infer the geographical location of a network entity based on its IP address. To show the effectiveness of our novel approach the evaluation is done in real-world infrastructures with specific focus on Europe. This area is seen to be more difficult for IP Geolocation than America, because of the unstructured and more complex infrastructure.

## 2. Scenario

The need for an accurate and reliable Geolocation service is illustrated by using the following real-world scenario. A sophisticated attack on a company network is detected by a behavior based Intrusion Detection System (IDS). The management of the company requires a clarification of the case. As part of a pre-forensic strategy, the real-world location of the attack source should be determined. Therefore, a Geolocation service is necessary, which uses the assumed IP address of the possible attacker and correlates delay pattern with the estimated distance. The extracted information are used to prove the evidence and to start further actions. Beside this, all data is recorded and stored with signatures to cope the forensic needs of law enforcement. Since the location of the aggressor is not known beforehand, a well distributed network of multiple Landmarks is required. The chosen Vantage Points have to be close to the measured target with respect to the infrastructure and the hop count. As closer such a reference host is to the target, the impact of interfering influences during the active latency measurement is reduced. Because of the high-performance connection, the ideal case are Landmarks placed within the backbone network topology. The accuracy of latency measurements and the optimal selection of Landmarks is directly related to the precise detection of a nearby target. We need to find a predefined number k center nodes with minimized maximum distance to the surrounding network topology. The amount of Landmarks needed is mainly influenced by the network load due to applied multiple measurements and cost-efficiency in terms of maintaining a probing infrastructure. This optimization problem is NP-hard and is based on the classical k-center problem for clustering. Furthermore, a stable model for Geolocation is necessary to reach reliable results at varying measurement values.

## 3. Related Work

Strategies for IP Geolocation can be classified in either IP mapping or measurement-based strategies [5]. IP mapping based approaches can be considered as passive by means of not interacting with the target. Instead, queries have to be conducted against public available Regional Internet Registry (RIR), the Domain Name System (DNS) LOC resource record, WHOIS-lookups, or commercial databases offered by geo-services [1, 5-7].

In comparison, measurement-based methods are relying on an active interaction with the target system, mostly done by probing the corresponding IP from Landmarks and recording the amount of time it takes for the response. It is based on the well-established assumption of an existent correlation between network latency and geographical distance, the location of the IP is inferred [8].

Since passive methods provide only rough and incomplete information [5, 6, 9], we focus on active measurement-based approaches, which has been proven to be more precise [10].

Measurement-based strategies can further be classified in simple delay measurements, e.g. *Shortest Ping* [4] or *GeoPing* [1], and constrained- as well as topology-based approaches. *Constraint-Based Geolocation* (CBG) [11] infers the geographic location of an IP address by using multilateration with distance constraints, hence establishing a continuous space of solution instead of a discrete one. The enhanced version *Topology-Based Geolocation* (TBG or Advanced CBG - ACBG) [4] additionally estimates the location of intermediate routers.

A combination of the latter ones with IP mapping is considered as hybrid strategy. Indeed, basically relying on active probing, hybrid approaches try to improve and verify the results [5].

The current "State-of-the-Art" in terms of active measurements are *Octant* [2, 12], *POSIT* [13], and *Spotter* [6]. They accumulate possibilities of the delays, hops, and distances to estimate the target region, whereas no exact location is defined. Thus, those models can be considered as not precise enough. Additionally these approaches do not take real-world circumstances from network into account. The most accurate results obtained by active measurement are up to 600 meters close to the target [10]. The proposed techniques has only been evaluated in research environments with a simple and homogeneous infrastructure or under assumptions like each company hosts their own webserver in-house. Such methods cannot be considered as generally valid for public networks.

Measurement-based approaches rely on Landmarks for active probing. They all have to face the dilemma of using as much as necessary but as few as possible Landmarks, known as *Landmark Problem*. The problem of selecting and positioning of those Vantage Points exacerbates this dilemma. In current research work almost no comprehensive information about how to deal with these problems is provided. In addition, usually Euclidean distances are used for the location estimation process, resulting in imprecise modelling. Ziviani et. al [8] provides an algorithm for placement of Landmarks. But their proposed integer linear programming (ILP) model obtains results with a fitness value up to two times of the optimal placement due to relaxation and integrity gap. Thus, it is not suitable for realistic and large scale scenarios and limited to specific areas only.

## 4. Model of Geolocation

Our proposed model for Geolocation uses a pre-defined number of Landmarks to actively probe the target IP address. The Landmark locations are optimized in respect to the surrounding network topology. The Round-Trip-Time (RTT) and the traced hop count are measured from multiple Landmarks to the target with different protocols and parameters. The data is used for a temporally correlation to estimate the distance to the geographical location of the target and further calculations. Through multilateration by using these results, the geographical location is inferred.

Furthermore, we identify the network device one hop before the target in accordance with tracerouting, since it can be assumed that the last hop previous to the target is usually located nearby [1]. In this way and by use of different protocols, we get a comparable value and try to minimize the influences of local firewalls and other unusual behaviors. Furthermore, special network configurations like proxies, VPN, MPLS Tunnels or anonymisation techniques can be identified and handled accordingly. If the target IP is not responding or traceable, we use the last identified hop for further calculation instead.

The conversion of the latency to a distance is calculated with support of trained and approximated logarithmic curves. The direct location is inferred by multilateration with support of different Landmarks measuring from several directions. Besides, we apply automatic filtering to select the correct measuring data and preliminary results. In the following sections, we present a complete and evaluated concept for IP Geolocation including Landmark selection, latency to distance calculation and target determination. An overview of the entire process model is presented in Figure 2.

## 4.1. Landmark Selection

The first problem is the identification and placement of Landmarks for probing a target in an infrastructure (see Figure 2, purple box). Since the latency-distance correlation depends on the region and their connectivity as well as on the used network components and materials, determining reasonable Landmark positions is an important challenge for a highly accurate IP localization service. For the selection of measuring points, the topology of the core network has to be known beforehand, e.g. by the ISP provider or public sources [14]. Through the large-scale deployment the selection of Landmarks should be limited to the high-performance backbone network (see Figure 3). Nevertheless, the Landmarks have to be identified in such an infrastructure with respect to minimize the maximum distance to the surrounding network topology of a single Landmark. With an improved selection, a single Landmark is closer to the target. As nearer they are, the distance according to the overall infrastructure is lower and the variance of measurement results is reduced due to the minimized amount of interfering sources. Furthermore, the selection of measuring points is limited to a given set of possible locations in the used public infrastructure. The optimized and preferred locations are mapped to the nearby possible Landmark positions. As a dense network of possible locations is available, the discrepancy is negligible.

As the target location is not known beforehand, a distributed network of Landmarks is highly recommended. An uniform distribution of Landmarks would be desirable to measure from different directions using different network paths. These optimized locations for Landmarks represent central nodes in a given network topology.

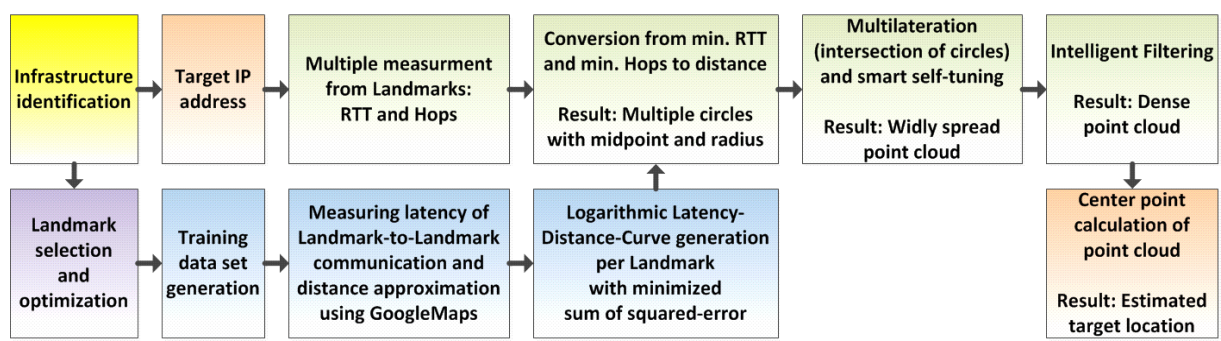

Figure 2. Process Model for our Geolocation service.

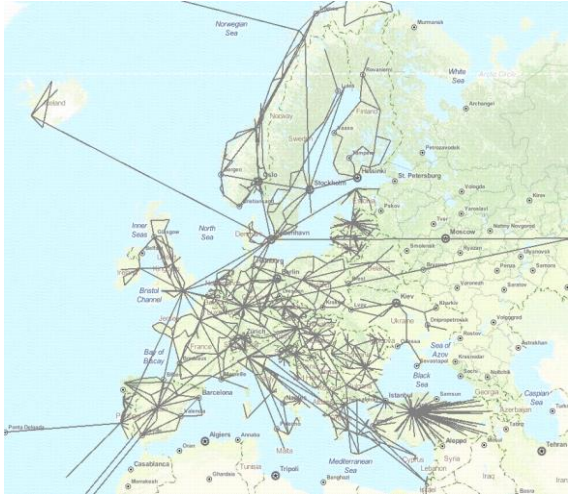

Figure 3. Internet backbone network topology in Europe [14].

We designed an adaptable algorithm to find optimized locations for Landmarks, called Dragoon (Diversification Rectifies Advanced Greedy Overdetermined Optimization N-Dimensions). It can also be used in the research field of data mining, clustering, facility location, and mirror server placement. The selected locations by Dragoon represent central nodes in a given topology.

As first step of initialization, an orientation mark is placed at the optimal center position of the given network topology according to place one measurement point. This mark is only for orientation to place the first Landmark afterwards and to obtain uniform distributed nodes for initialization. The first Landmark is placed at the position of a network node which is farthest away from the orientation mark. Subsequently, the remaining number of the pre-defined amount of Landmarks is placed using the 2-Approx strategy [15]. It calculates for every network node the distance to all placed Landmarks and chooses the node with the largest distance to their closest Landmark as the new location to place the next Landmark. The orientation mark is only used for the placement decision of the first Landmark with the 2-Approx strategy, afterwards it is removed. Thereby, we obtain a specific solution of the 2-Approx placement strategy, which normally places the first node randomly.

After the initialization, the algorithm starts with the iterative refinement rule to optimize the locations and find the final list of Landmarks. It checks all possible locations around the current position of a Landmark and tests all connected nodes with a direct edge to the current position (see Figure 4). If the new location improves the overall situation, the algorithm replaces the current position with the new one. This is done with respect to the specified optimization criterion. In our case, it is the maximum distance counted by hops with Dijkstra's algorithm. If this major optimization criterion is unchanged, the algorithm will use an additional criterion to decide between equal solutions. We use an average or mean criterion to identify a minor improvement. In each iteration step, all network nodes of the observed infrastructure are (re)assigned to their closest Landmark. Every Landmark is allowed to shift its position only once. This leads to a stepwise improvement and avoids a too fast stagnation in a local optimum. The Landmarks are selected in the same order as they are added to the network topology.

This iterative optimization is repeated until all Landmark positions do not change any more. The algorithm accepts only improved positions in every step. Therefore, the 2-approximable condition holds and it will always terminate. The selected positions are used as Landmarks to send packets to the target and measure inter Landmark communication.

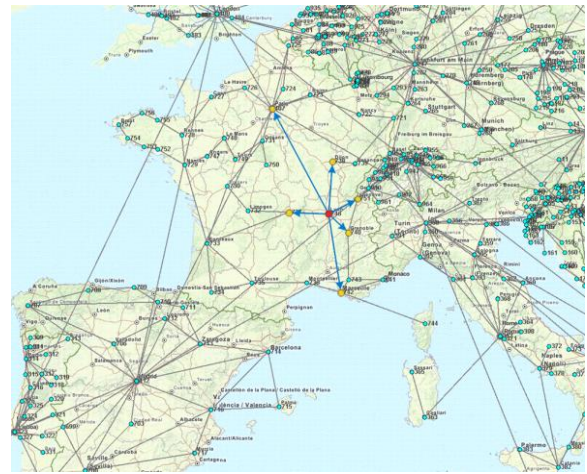

Figure 4. Iterative optimization step of the algorithm Dragoon.

### 4.2. Delay Analysis of Measurement Results

The determined Landmarks trace and ping the target several times to measure the RTT, delay variance and hop count. Based on the theoretical maximum speed of about 225,000 km/s that packets can travel in networks cables, we need to measure the time in micro seconds to limit the error in transformation to distances below 225 meter.

The delay between two network entities consists of a deterministic and a stochastic delay. The deterministic delay is composed by the minimum processing time of each router, the transmission delay, queuing delay, and the propagation delay. It is fixed for any given path in relation to the distance based on the hardware and connection conditions. The deterministic delay represents the minimum transmission delay, which we are interested in. These components are considered in our concept described in Section 4.3. The stochastic delay composes the queuing delay and the variable processing time as well as buffering at each intermediate router that exceeds the minimum processing time. For accurate results, we need to avoid such effects and keep the stochastic delay minimal. To counter the stochastic delay, several measurements are necessary to get a value close to the theoretical minimal RTT. For our evaluation, we send ten requests from each Landmark to the target. The measurements are done with specialized parameter to improve the delay, see Section 4.8.

Furthermore, the effect of stochastic influences is respected by our model through a dynamic and self-optimized *LC* factor, measurements from multiple Landmarks, and transformation from delay to distance including a minimal, average overhead.

## 4.3. Conversion of Measurement Results to Distance

Another problem is the calculation of the distance out of the latency (see Figure 2, blue boxes). As there is no concrete mathematical model for this application area, we use a reference function to map the measured latency to a distance. As we know from previous work of Center for Applied Internet Data Analysis (CAIDA) [7], the correlation between latency and real distance follows approximately a logarithmic curve:

$$distance = p * ln( q * latency + n ) + m.$$

An example of such an individual Latency-Distance-Curve is presented in Figure 5. The parameter p, q, n and m for such a fully parametrized curve are unknown and need to be calculated with a training data set. *Latency* is *RTT / 2* subtracted by the *average delay* of *hops* including a stochastic part, which is about *0.055 ms* per hop [16]. An additional pre-defined processing delay of about *0.11 ms* is subtracted once to reflect the last router generating the response packet [8, 17]. The logarithmic correlation compensates the large transmission delay through the processing units compared to the signal propagation speed in the conductor and smooth stochastic delay. This influence is particularly strong at the "Last Mile" connection to end user with cheap network nodes. In comparison, most current research work abstract this correlation as a linear function. Such modelling do not take the different TIER network levels into account.

For the curve reconstruction, we approximate the function using curve fitting with a minimized sum of squared-error. The training data is based on multiple RTT measurements between a Landmark and several known locations, mostly Landmark-to-Landmark. As the entire network topology is complex and usually unknown, the real path lengths has to be estimated. With the knowledge that cables are typical installed along roads through practical construction, the road network can be used. The geographical distances in our model are calculated with the service of Google Maps. A path especially along the network infrastructure is obvious longer than a straight or orthodromic line on the earth surface. Hence, underestimated real distances would lead to wrong estimated target locations as used by all other related work in Section 3. A comparison of our approach shows the impact of the underlying distance base in contrast to the WGS84 reference ellipsoid as well as orthodromic distances, also known as great-circle. Thereby, we achieve a much higher accuracy than modelling the earth as a ball without mountains and valleys, rotation flattening effect and applied Euclidean distances. Only if the service of GoogleMaps do not provide a realistic distance, we have to use the orthodromic distance with added 10% as fallback estimation.

A training data set of latency and distances is used to train our model and calculation of the individual Latency-Distance-Curves of every Landmark. These are highly important to transform the measured delays to a target in distances for the location estimation. The training data consists of measured delays between known locations, mostly Landmark-to-Landmark communication and distances calculated by Google Maps.

Every Landmark has a unique location in the network topology including the surrounding connections, network devices and terrain. So the Latency-Distance-Curve is Landmark specific. As we select Landmarks in the backbone topology with high-performance connection, the resulting curve will be in general overestimated. We counter this by applying an adaptable factor for all Landmarks, called *LC*. This factor allows us a general shifting of the logarithmic curves to decrease the estimated distances. The factor should be smaller than 1 and enables our model to be adaptable. The automatic adjustment and training of the logarithmic curves is achieved to improve the estimated target location. The process of smart self-tuning is explained in Section 4.6.

In general, the correlation is expected to follow a direct linear relation, because the network behavior does not include non-monotonic components. In this context, the curvature of the curve reflects partially the inaccuracy of a model and counter the aforementioned effects.

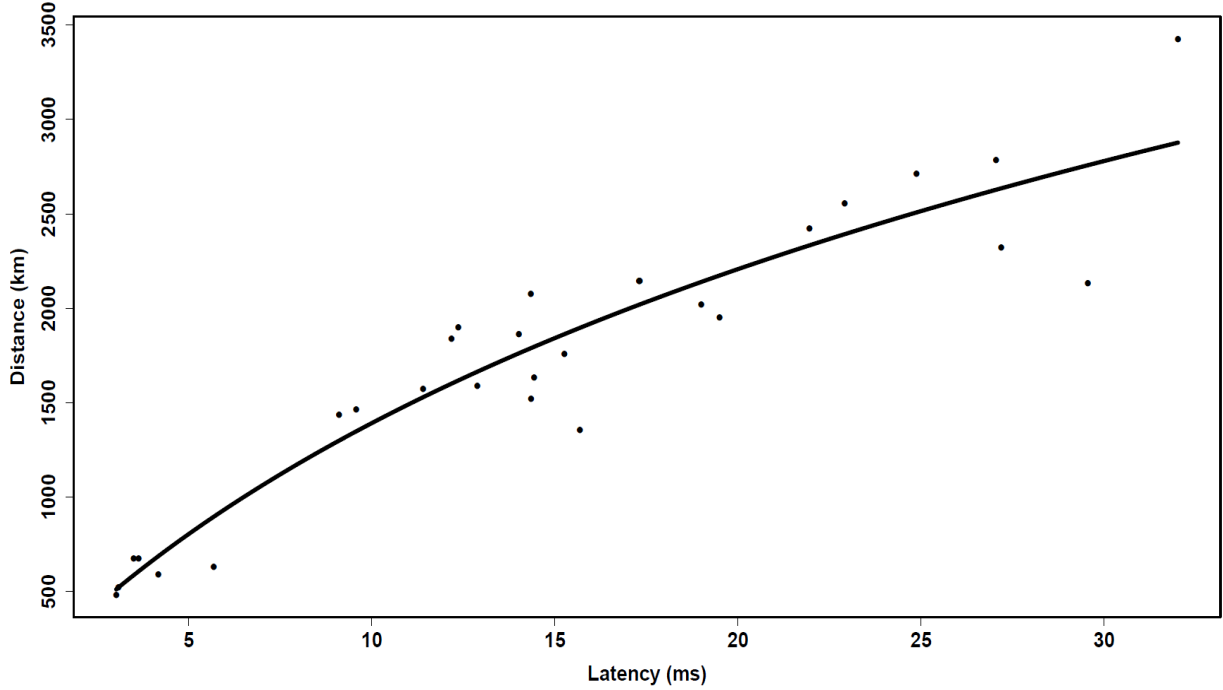

Figure 5. Example of an individual Latency-Distance-Curve of a single Landmark.

## 4.4. Target Localization by Lateration and Adapted Data Transformation

The following steps of the target localization process are illustrated in Figure 2 (green boxes).

To determine a location, a selection of Landmarks is probing the target by ping and traceroute with different protocols enabling to bypass simple firewall restrictions. A geographical location of a target is determined using the measurement method of lateration. It uses two known geographical Landmark locations and the estimated distance from each to the target. For every Landmark, we obtain a circle with known midpoint ($x_{LM}$, $y_{LM}$) and radius $r_{LM}$ representing the minimal delay from there to the target:

$$(x_{Target} - x_{LM})^2 + (y_{Target} - y_{LM})^2 = r_{LM}.$$

We calculate the intersection of two circles by equation method to determine the target location. To increase the precision, multiple Landmarks are used for probing the targets IP address and calculating the lateration. As we are dealing with distances by lateration, the received data of RTT and hop count are converted to a distance with support of the individual Latency-Distance-Curves. Thus, the location of each Landmark and the distance from there to the target are known. As the radius is in km and the midpoint in geographical degree, both are transformed in a common unit of degree in the 2-dimensional Euclidean space for lateration. One degree represents 113.325 km at the equator for latitude and longitude. Along the longitude from the equator to the poles, the spacing is reduced. By dividing the radius through it, we obtain equal spacing in degree.

## 4.5. Different Cases of Intersection During Lateration

As result of the lateration process, four different situations may occur and have to be interpreted in context of Geolocation. Figure 6 illustrates the case that the circle areas do not overlap and that there is no intersection. The target location is estimated in between the space of two Landmarks in relation to the circle radius. Therefore, we enlarge the radius of both circles simultaneously with a constant factor.

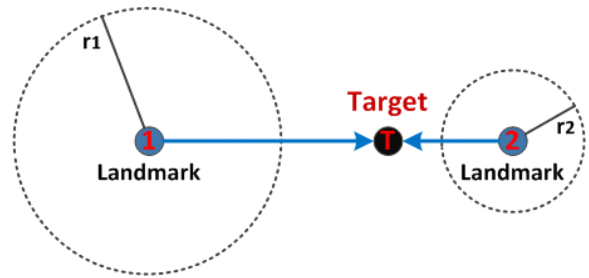

Figure 6. Zero intersections. Non-overlapping circles.

If the circles are completely overlapping (see Figure 7), the target location is likely to be on the circle site, where the circle rings are closest together. To determine the intersection, we reduce the larger radius.

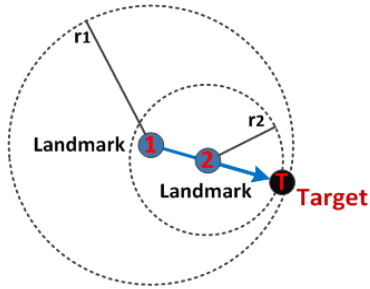

Figure 7. Zero intersections. Overlapping circles.

Figure 8 illustrates the ideal case, where both circles intersect in only one point, which represents the assumed location of the target. For this case, it is necessary that all three nodes are positioned on a straight line, which is not often.

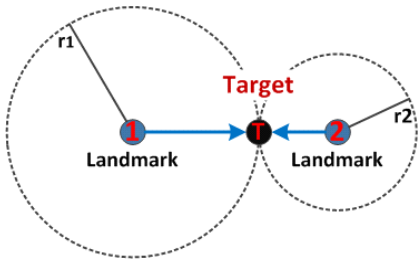

Figure 8. One intersection point.

The last possible case is presented in Figure 9. It shows partly overlapping circles creating two intersection points. The searched location is estimated on one of these two points. As we deal with probabilities, we accept both points as possible target locations. In this stage, we do not take further information from other Landmarks into account to decide between the two points, because these information are also not verified. In the subsequent evaluation of the intersection points it is expected that more points are dense clustered at the right location (see Figure 10 and 11). Thus, the wrongly estimated target locations are filtered out.

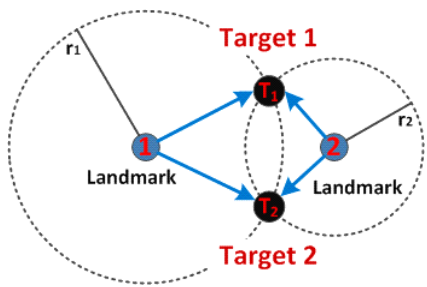

Figure 9. Two intersection points.

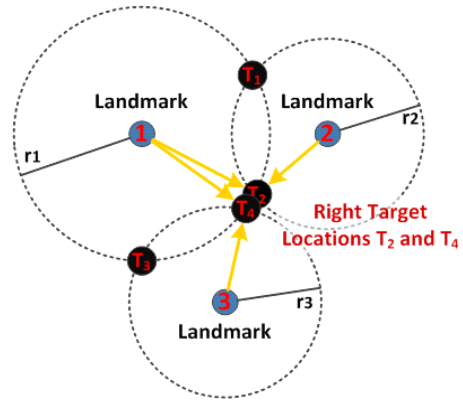

Figure 10. Lateration with multiple intersection points.

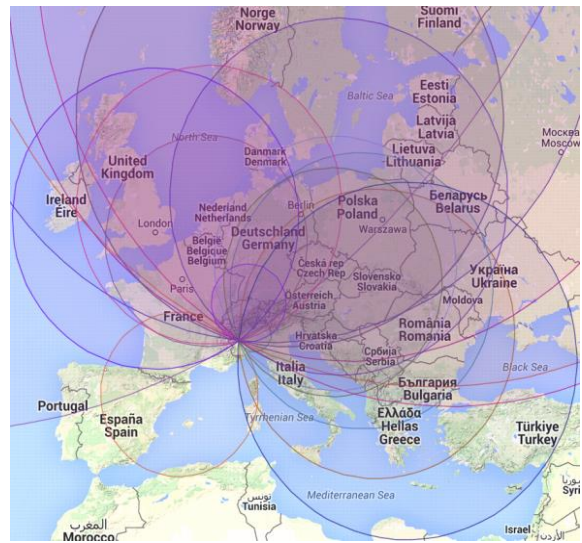

Figure 11. Example Visualization on a target for location determination.

### 4.6. Smart Self-tuning

As result of the conducted multilateration, we obtain a point cloud with estimated target locations (see Figure 12). If the measurement results, including conversion, are leading to a correct distance to the target destination, many intersection points are at the same location (see Figures 10 and 11). By measuring variations, the circle radii of all Landmarks vary in approximately the same level, leading to inaccurate intersections and overestimated distances. In order to counter the current network utilization and measurement deviations, an automatic optimization is carried out via the *LC* factor.

We calculate from all intersection points the orthodromic distance to all other intersection points. Afterwards, the distances are sorted beginning from the shortest. All distances smaller than the median value are summed up. The 50% originates from the maximum of two intersection points of two circles, whereas only one point can be the estimated target location. The median value represents the deviation of the temporary multilateration results. With a decreased value, we expect that more intersection points are shifting to nearly the same location and the point cluster at the real target location becomes more dense. In an iterative process, we decrease the radii of all Landmarks by 1% in each iteration step as long as the sum decreases.

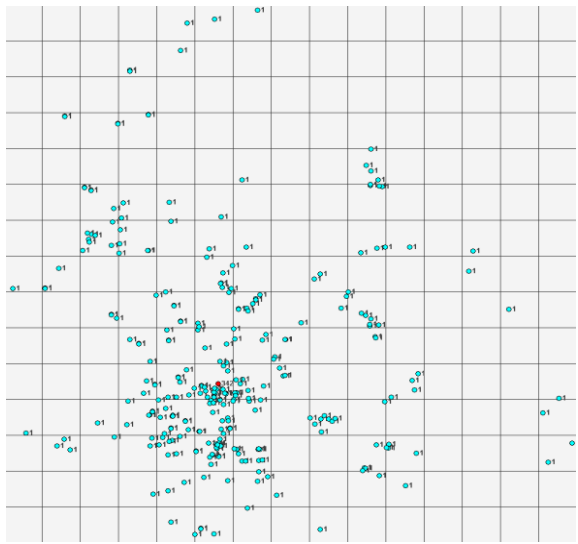

Figure 12. Cloud of intersection points (blue) with a dense area close to the calculated center node (red).

### 4.7. Intelligent Filtering of Correlated Information and Target Localization

From the mentioned prior steps, we obtain a cloud of multiple points, each representing a certain probability to be the target location. After applying the smart self-tuning process, the cloud should include a dense cluster. It represents the area with a high probability to be the real location. The problem is the identification of it. Therefore, we start an iterative filtering procedure to narrow down the target location. We filter outliers and misleading points. With our Dragoon algorithm, adapted to a constrained free center placement, we calculate the center location of all points in the cloud. Dragoon optimizes the location of a single center according to the minimal average distance using an orthodromic metric. Our algorithm iterative tests all points on a discrete grid with a defined distance $e$. If one of the tested locations results in a better performance, this location is used for the next iteration step. If no location leads to an improvement in a step, we successively decrease the granularity of the grid ($e_{new} := e_{old}/2$). This process is repeated until the grid distance $e$ is smaller than the maximal accepted deviation. It is necessary to define a limit for the maximal deviation to terminate the optimization process. The processing steps of the iterative optimization are illustrated in Figure 13. The left side illustrates the movement to an improved spot. The right side shows the increased granularity of the grid by bisection.

Affiliating, all points with largest distances to the calculated center location are ignored in further steps. This process of center placement and filtering is iteratively repeated until the amount of points left in the cloud is below the number of used Landmarks, resulting in one location for each Landmark. These points represent the searched, dense clustered region to expect the target location to be likely in the center of this area. Once more, we calculate the center point of these points, which reflects our estimated target. The distance between the determined center location and the farthermost point in the dense clustered region is seen as the expected error deviation.

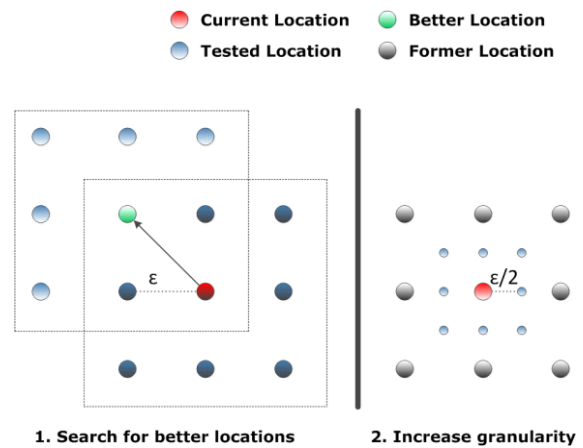

Figure 13. Iterative optimization stage of the algorithm Dragoon by free placement constraint.

### 4.8. Measurement Specification

Constant network paths are important for measuring and training of delay overheads. Considering multiple measurement results, we are only interested in the minimal RTT in order to obtain the correct distance. In addition, the probing packets are marked with the traffic classification low-latency in differentiated services (DiffServ) field, which supports the measurement to get minimal RTT. This is supported by small packet sizes, which avoids bandwidth rules for large packets or high utilization rerouting rules. Apart from the delay measurements, the hop count on the path has to be determined. We use different ISO/OSI Layer 3+ Protocols to detect network nodes, like ICMP, UDP, and TCP, because router and firewalls react differently dependent on it [1]. Before the raw measurement results are used for lateration, these are specially selected. The different traces are filtered after minimal hop count per protocol representing the minimal path length. Every hop causes a probabilistic delay through processing, which we can only be estimated. Thus, our focus is to keep the impact low. Of each used protocol, we select the maximum value to obtain the truthful number of network nodes on the path. The small time and hop deviation of asymmetric connections are balanced by using minimal RTT and the adaptive model construction with predefined, static delay overheads and the self-tuning *LC* factor. The minimum delay and hop deviation in comparison to the real path is balanced with the adaptive model construction and self-tuning *LC* factor.

## 5. Evaluation and Assessment

In order to verify our concept and algorithm, we set up multiple experiments. As our focus is Europe including private users with moderate internet connections via DSL, we use public available information and research data from the TIER 1 topology like [14]. We have to build a set of distributed reference hosts to which we have access to. For this purpose we are using the RIPE Atlas Project [18] providing us with over 8200 possible well-known nodes for probing. RIPE Atlas provides a comprehensive distributing of nodes in terms of different bandwidth and population density. Thus, for Europe we are not limited to residential areas only.
The first step is to calculate optimal positions in respect to the given topology. Afterwards we compare these locations to our set of reference hosts to find direct matches or nearby nodes according to latitude and longitude. The next steps follow exactly our presented model in Section 3. To determine the hop count and the RTT we use Traceroute, Paris Traceroute, and ICMP echo request provided by the RIPE Atlas measurement interface. By using the hop count and the measured minimum delay out of ten measurements, a logarithmic curve is calculated in order to represent a correlation between measured latency and road network distance. The Latency-Distance-Curve reconstruction is calculated parameter pairwise iteratively with the tool R and the curve fitting method *nls*. After probing the target IP address from each Landmark, the curve is used to convert the RTT and hop count to a geographic distance. Using the calculated distance as well as the knowledge of longitude and latitude of each probing Landmark, Dragoon is able to infer the location of the target.

**Result 1:** For a suitable measuring method, we evaluate the influence of the packet size on the latency. Figure 14 compare the latency of the minimum packet size with the almost maximum packet size avoiding packet fragmentation. The time period is about 30 days whereas every measurement point represent the average delay within one hour for a large sample set of communications in Europe, using the minimum delay for each communication. The influence is negligible as long as the no packet fragmentation is mandatory.

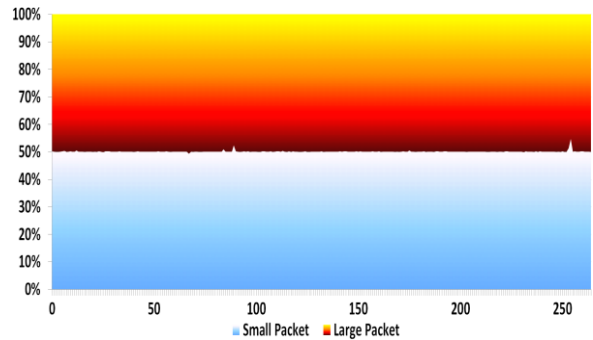

Figure 14. Delay difference between small and large packets.

**Result 2:** For an improved target estimation, we analyse the average latency distribution during a long time horizon. Such an aspect influences the Latency-Distance-Curve and the calculated distances based on the latency. Figure 15 shows the latency distribution for 30 days, whereas every measurement point represent the average delay within one hour, using the minimum delay for each communication. The delay is almost the same and independent from day time as well as week days. There are some outliers which are hard to predict. Especially the red marked pillars results from a Microsoft Security Bulletin and Patch Day. To counter such effects, the self-tuning *LC* factor is necessary. All measurements should be applied in a short period of time to have similar conditions and to obtain a suitable reference RTT.

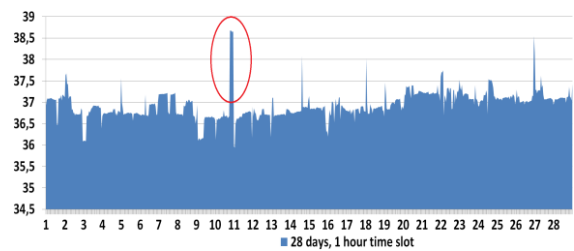

Figure 15. Average delay over 28 days between Landmarks.

**Result 3:** The Table 1 shows an excerpt from the comparison of estimated target locations obtained by different applied Landmarks, which are identified by Dragoon and 2-Approx. Since the used scenario is too large for common ILP solver, we used the alternative algorithm 2-Approx. It illustrates the impact of the selected Landmarks to the results of IP Geolocation.

The comparison of the Latency-Distance-Curve based on GoogleMaps and orthodromic distance shows the difference in the curvature. The function using orthodromic distances is more curved and visualizes the impact of the "Last Mile"', which is to a limited extent covered by the modelling. For this reason and because of improved and more stable location estimation, the further calculations are based on the curve using Google Maps.

As we know that the logarithmic curve is too optimistic, we first identify a uniform *LC* factor to counter it and to analyze the model. For the curve based on Landmarks optimized by Dragoon, the *LC* factor is about *0.7* for distances obtained by Google Maps (GM) and Orthodroms (OD). This means the distance to a given Latency is 30% to large. For the function based on Landmarks identified by the reference algorithms, the optimized *LC* factor is *0.9*. With an self-optimized factor per target, we are able to achieve more precise results (see Table 2). This justifies our assumption that the Landmark selection and the reference distances have a strong influence on the result.

Table 1. Example comparison of the derivation between the location estimation to the real geographic location using different Landmark sets.

| Target | Dragoon (GM) | Dragoon (OD) | Reference Algorithm |
|---|---|---|---|
| 1 | 120 km | 117 km | 350 km |
| 2 | 136 km | 536 km | 1600 km |
| 3 | 221 km | 108 km | 113 km |

**Result 4:** The influence of the amount of Landmarks on the accuracy of the results is analyzed by two different data sets. We determined 17 target locations with the amount of 20 and 30 Landmarks, whereas the Landmark positions are independently optimized with our algorithm Dragoon. Table 2 compares the accuracy of the target determination for several IP addresses and their location. It shows the precision of our approach. The accuracy does not increase with the amount of Landmarks. Even if the results get more stable in relation to the variation of the parameter in our model, it is more important to reach accurate measurement results. Also the selection and accumulation of the right input data has a highly impact on the target estimation. Filtering through the model is only to counter measurement fluctuation. With a higher amount of Landmarks, the point cloud contains more possible locations, which are tougher to filter and to detect outliers. About 41% of the targets are localized within 50 km. More than the half of the results have a deviation of less than 100 km for only 20 Landmarks covering entire Europe.

**Result 5:** As other publications does not respect the special case of overlapping circles during lateration like illustrated in Figure 7, we evaluate the influence of this aspect in our model and the deviation of the determined target locations. Ignoring overlapping circles result in about 11% higher deviation as our modeling approach. Also the abstraction to the midpoint of the inner circle result in about 7% higher inaccuracy.

**Result 6:** Since active measurements have always been considered to be more precise as passive IP mapping-based techniques, we compared our approach to the passive services of MaxMinds [19] and WhoIs [5] as well as the active techniques Spotter [6] and TULIP [20] of the Stanford University (CBG, ACBG, Geoplugin and FreeGeoIP). This measurement shows the precision of our entire concept in relation to other state of the art techniques for example [13, 21].

As shown in Table 3. Dragoon is outperforming Whois by 67.21% with over 275km less average deviation. Interesting is, that MaxMind performs even worse than Whois with 70.96% and over 328.36 km average deviation. Also the other active measurement techniques show a large deviation for our targets. Since their approaches are not available for a direct comparison for single targets, we use the online service TULIP. It provides different re-implementations of these state of the art techniques and queries geodatabases. In contrast, our active approach respects the fluctuation of network delays as mentioned in result 2. It uses a more accurate estimation of network distances by road networks and improved placement of Landmarks. Additionally, the new model takes the hop count of the network path into account as well as further aspects as described in Section 4.

Table 2. Comparison of the derivation in km between the location estimation to the real geographic location using different amounts of Landmarks.

| Target | 1 | 2 | 3 | 4 | 5 | 6 | 7 | 8 | 9 | 10 | 11 | 12 | 13 | 14 | 15 | 16 | 17 | Avg. |
|---|---|---|---|---|---|---|---|---|---|---|---|---|---|---|---|---|---|---|
| 20 LM | 0 | 3 | 11 | 23 | 28 | 38 | 43 | 87 | 95 | 148 | 176 | 194 | 205 | 268 | 268 | 307 | 390 | 134 |
| 30 LM | 214 | 77 | 198 | 87 | 169 | 70 | 165 | 143 | 353 | 135 | 198 | 243 | 207 | 155 | 107 | 222 | 277 | 178 |

Table 3. Comparison of Dragoon with 20 Landmarks to active (non-italic) and passive (italic) IP Geolocation services.

| Target | Dragoon | *WhoIs* | *MaxMind* | Spotter | CBG | ACBG | Geoplugin | FreeGeoIP |
|---:|---:|---:|---:|---:|---:|---:|---:|---:|
| 1 | 0 | *14* | *1* | 478 | 481 | 481 | 1 | 1 |
| 2 | 3 | *24* | *2* | 193 | 179 | 179 | 2 | 2 |
| 3 | 11 | *1* | *64* | 221 | 109 | 109 | 64 | 64 |
| 4 | 23 | *14* | *13* | 738 | 806 | 795 | 13 | 22 |
| 5 | 28 | *15* | *2* | 6765 | 6765 | 6765 | 6765 | 6765 |
| 6 | 38 | *12* | *2* | 335 | 459 | 459 | 2 | 2 |
| 7 | 43 | *379* | *370* | 136 | 101 | 101 | 366 | 366 |
| 8 | 87 | *453* | *381* | 567 | 676 | 771 | 382 | 382 |
| 9 | 95 | *16* | *65* | 495 | 205 | 205 | 65 | 65 |
| 10 | 148 | *24* | *3* | 475 | 586 | 586 | 3 | 3 |
| 11 | 176 | *433* | *477* | 411 | 483 | 483 | 485 | 485 |
| 12 | 194 | *69* | *402* | 951 | 1063 | 1015 | 396 | 396 |
| 13 | 205 | *5481* | *5474* | 477 | 172 | 163 | 5390 | 5389 |
| 14 | 268 | *2* | *341* | 401 | 866 | 866 | 347 | 347 |
| 15 | 268 | *1* | *13* | 46 | 46 | 46 | 13 | 6 |
| 16 | 307 | *28* | *254* | 11 | 64 | 64 | 6 | 254 |
| 17 | 390 | *0* | *1* | 111 | 4 | 4 | 1 | 1 |
| **Avg. Deviation** | **134** | **410** | **463** | **754** | **768** | **770** | **841** | **856** |

## 6. Limitations and Discussion

If the rough underlying infrastructure is unknown, a determination of Center Nodes is not possible. Also helix like infrastructures will result in suboptimal calculations, hence more coarse-grained location estimations. Our approach obtains only precise results for targets located in the bounding box created by the Landmarks. Since it has to be assumed that at least the TIER 3 infrastructure provides in comparison to TIER 1 less capacity and connectivity, the logarithmic curve based on the TIER 1 infrastructure has to be considered optimistic. To overcome these shortcomings a more comprehensive knowledge of the TIER 1 to 3 infrastructure, the "Last Mile" as well as the transmission medium and the network load of different components is needed. Thus, the amount of Landmarks needed for probing is highly depending on that knowledge. Further factors which may have influence on the location estimation are the used protocol for probing. Traceroutes can be done by different protocols and algorithms, hence causing more or less overhead while different processing steps on the packet path. Caused by its design IPv6 may have impact on the measurements results, too. For moderate connected areas the knowledge of the used transmission technique and medium is even more important, as directional radio, satellite connections and for instance Googles Loon has to be handled differently in determining the real path length. In these cases the use of the road network might not be a reasonable solution for this task.

## 7. Conclusion

In this paper we propose an advanced model for IP Geolocation based on active measurements. We show the general usability of time measurements for high precision based on the selection and position of Landmarks. Our model is able to identify enhanced Landmark locations for probing. In addition, it compensates influences of real-world environments by a self-optimizing process. We have shown that the accuracy of the location determination does not increase with the amount of Landmarks used for probing, but the results are less vulnerable to measurement fluctuations. Considering other publication do not respect special cases like overlapping circles during the process of lateration, our approach performs better. Also, the usage of road networks for distance estimation and respecting hop count leads to improved results, except for mentioned limitations. It localize more than 40% of the targets with a deviation less than 50 km. Currently we are evaluating our model in respect to moderated connected areas as well as intercontinental connections. We still have open issues in respect to model enhancements by possible overheads caused by IPv6 and further optimizations like analyzing the influence of the tracing algorithm on our model. Assigning every intersection point an individual probability may increase the identification of the dense cluster and therefore the target determination. Dynamic selected Landmarks, which are located near the estimated target, lead to improved measurement values.

## 11. Acknowledgements

We want to thank Sebastian Seeber for providing the needed credits for applying the measurements using RIPE Atlas. Additional thanks for supporting us goes to RUAG Schweiz AG, Division - RUAG Defence.

This work was partly funded by FLAMINGO, a Network of Excellence project (ICT-318488) supported by the European Commission under its Seventh Framework program.